# Straightforward Method to Orient Black Phosphorus from Bulk to Thin Layers Using a Standard Green Laser

**Authors**

Etienne Carré[1,2]*, Frédéric Fossard[1], Jean-Sébastien Mérot[1], Denis Boivin[3], Nicolas Horezan[3], Victor Zatko[4], Florian Godel[4], Bruno Dlubak[4], Marie-Blandine Martin[4], Pierre Seneor[4], Etienne Gaufres[5], Julien Barjon[2], Annick Loiseau[1], Ingrid Stenger[2]

[1]Université Paris-Saclay, ONERA, CNRS, Laboratoire d'Étude des Microstructures (LEM), F-92322 Châtillon, France
[2]Université Paris-Saclay, UVSQ, CNRS, Groupe d'Étude de la Matière Condensée (GEMaC), F-78035 Versailles, France
[3]Université Paris-Saclay, ONERA, DMAS, F-92322 Châtillon, France
[4]Laboratoire Albert Fert, CNRS, Thales, Université Paris-Saclay, F-91767 Palaiseau, France
[5]Laboratoire Photonique Numérique et Nanosciences, CNRS-Institut d'Optique-Université de Bordeaux, F-33400 Talence, France
* Corresponding author

**Abstract**

The crystallographic orientation of anisotropic 2D materials plays a crucial role in their physical properties and device performance. However, standard orientation techniques such as transmission electron microscopy (TEM) or X-ray diffraction (XRD) can be complex and less accessible for routine characterization. In this study, we investigate the orientation of black phosphorus (BP) from bulk crystals to thin layers using angle-resolved polarized Raman spectroscopy (ARPRS) with a single-wavelength (514 nm) Raman setup. By incorporating thickness-dependent interference effects and anisotropic optical indices, this approach provides a reliable framework for orientation determination across different BP thicknesses. The method is validated through direct orientation measurements using TEM and Electron Backscattering Diffraction (EBSD), confirming its applicability to both thick and ultrathin samples. Given its simplicity and compatibility with widely available Raman setups, this approach offers a practical solution for characterizing BP orientation without requiring advanced structural characterization techniques.

## I. Introduction

The field of two-dimensional (2D) materials has expanded rapidly since graphene's discovery in 2004, introducing novel properties that have reshaped many areas of physics and materials science. In contrast to graphene and other well-studied isotropic 2D materials—such as hexagonal boron nitride (hBN) and various transition metal dichalcogenides (e.g., $WS_2$, $MoS_2$)—certain 2D materials exhibit pronounced in-plane anisotropy[1]. Notable examples include transition metal trichalcogenides like $TiS_3$ or $As_2S_3$, as well as rhenium- ($ReS_2$, $ReSe_2$) and germanium-based (GeAs, GeSe, GeP) compounds, where electronic, optical, and mechanical responses vary significantly along different crystallographic axes [2]. This intrinsic anisotropy makes precise crystallographic orientation a critical factor for optimizing device performance, necessitating robust and accessible characterization techniques. Among these anisotropic 2D materials, black phosphorus (BP) stands out due to its tunable bandgap and high charge carrier mobility at room temperature[3,4], as well as its direct optical gap adjustable from the visible [5–12] to the mid-infrared spectrum[13–18]. These properties position BP as a promising bridge between graphene and transition metal dichalcogenides[19], with significant potential in applications such as infrared detectors[20,21], photovoltaics[22,23] and telecommunication[24–26]. Moreover, BP's low-symmetry structure results in directional variations— enhanced luminescence efficiency[10] and electronic transport[4,27] along the armchair axis, versus superior mechanical resistance[28] and thermal conductivity[29] along the zigzag axis—making the reliable identification of its crystallographic axes essential for device optimization[4,26,30,31].

Several methods have been proposed to identify BP's crystallographic orientation. Diffraction techniques like X-ray diffraction (XRD)[32] or transmission electron microscopy (TEM)[33,34] provide a direct identification of the crystalline orientation, with XRD suited for bulk samples and TEM for very thin samples on grids. However, these techniques are less practical for BP flakes exfoliated onto substrates, which are typically tens of microns wide and a few nanometers thick. Angle-Resolved Polarized Raman Spectroscopy (ARPRS) offers a non-destructive alternative that can be applied across a range of thicknesses[12,35]. ARPRS indirectly determines crystallographic orientation by probing the Raman tensor (matrix representation of phononic mode symmetry) through controlled polarization of incident and analyzed light[32,36]. However, challenges arose in early studies, particularly due to variations in BP structure descriptions (crystallographic abc axis[37,38], several cartesian yzx bases[4,10,29], inherited base from the zigzag, stacking and armchair representation in graphene and nanotubes[39,40]) and inconsistencies in Raman intensity maxima, initially attributed to conflicting zigzag or armchair axis assignments[34,35]. Moreover, Raman intensity is governed by a wide range of effects, both intrinsic (electron–phonon and electron–photon coupling, anisotropic optical absorption) and extrinsic (interference effects with the substrate, sample degradation, local heating under the laser)[41,42]. Consequently, parameters such as thickness[32,36,43], wavelength[34,36,44,45], or temperature[32] have all been shown to shift the angular position of the Raman intensity maxima, particularly under the widely used 514 nm excitation [33,36,46]. Multi wavelength investigations demonstrated that shorter wavelength like 442 nm (2.8 eV), induce resonant Raman scattering conditions that resolve ambiguities across BP samples[36,46,47]. However, these setups require specialized instrumentation not accessible to all laboratories, many of which operate with a single excitation wavelength (typically 514 or 532 nm)[48].

Recognizing the constraints of standard Raman setups, we propose a validated ARPRS-based protocol at 514 nm that ensures reliable BP orientation without requiring advanced instrumentation. To that end, our approach combines ARPRS measurements with direct orientation methods such as TEM or Electron Backscattering Diffraction (EBSD). Specifically, we first validate a method to distinguish (a) and (c) axes in thick BP crystals, optimizing tensor parameters through TEM verification. We then analyze thickness-dependent behavior using theoretical and experimental data, comparing ARPRS with EBSD for thin layers to clarify previous discrepancies. The systematic approach presented here provides a robust protocol for the crystallographic orientation of BP flakes, implementable with standard Raman equipment. This work is expected to facilitate BP orientation in varied research settings, particularly for applications in device fabrication where anisotropic BP properties can be strategically leveraged.

## II. Materials and methods

*Sample fabrication and characterizations*

The BP crystal has been stored in glove box under argon atmosphere (<0.5 ppm $O_2$, <1 ppm $H_2O$) to prevent its photo-oxidation under ambient conditions. BP thin layers are also exfoliated in glove box with conventional PDMS (1:10 ; cross-linked 1h at 80° C) and then deposited on a $SiO_2$(300 nm)/Si substrate. A hermetically sealed suitcase filled with argon gas was utilized to transport and introduce BP samples into an atomic layer deposition (ALD) growth chamber (BENEQ TFS 200) for the purpose of passivation. A continuous 1nm thick layer of $Al_2O_3$ was deposited on the BP samples through the reaction of tri-methyl-aluminum and $H_2O$, which provides an effective barrier against air exposure[49,50]. The flakes studied were then observed with an optical microscope and their thickness measured with an AFM (Bruker Innova AFM), both equipments being implemented in a glove box to avoid contamination by oxygen and water.

*Raman spectroscopy experiments*

Raman measurements were performed on a LabRAM HR800 Horiba Jobin Yvon Raman spectrometer equipped with an Argon laser (514 nm) with a 100x objective (NA=0.8) for an illumination spot diameter of about 1 µm. The beam is then analyzed with a monochromator equipped with a grating of 1800 lines/mm and a CCD detector cooled with liquid nitrogen. The laser power was kept below 2 mW and an argon gas flux was kept flowing over the sample surface to avoid sample degradation. The polarization at the laser output is linear and vertically oriented. Its rotation is ensured by means of a half-wave plate. The analyzer is located at the entrance of the spectrometer. The half-wave plate and the analyzer are crimped in motorized rotating mounts controlled by computer. The diffraction grating of the monochromator being more sensitive in one direction of polarization, we associate a quarter-wave plate to the analyzer whose neutral axes are set at 45° of its axis of passage.

*Electron Microscopies*

*FIB* - 60 nm TEM lamellas were fabricated using a FEI Helios 660 dual-beam microscope that combines a focused ion beam (FIB) column operating at 30 kV with a field emission gun (FEG) scanning electron microscope (SEM). A two-steps platinum film deposition is utilized as a protective layer for the BP against the $Ga^+$ ion beam, which is responsible for ejecting atoms from the sample and machining the material at a nanometric scale.

*TEM* - TEM slides were examined in a Zeiss Libra 200 MC equipped with an electrostatic CEOS monochromator, an in-column Ω filter and a Gatan ultrascan 1000 CCD camera. We used the High Resolution phase contrast imaging mode (HRTEM) at 200 keV providing a resolution limit below 150 pm.

*SEM-EBSD* – EBSD measurements were performed in a Zeiss Merlin field effect gun scanning electron microscope equipped with a Nordiff EBSD diffraction camera and processed with the TSL Oim software. The measurements were performed at an accelerating voltage of 20 kV.

*ARPRS model for anisotropic BP*

The incident field is calculated at a depth, labelled y, in the black phosphorus layer, considering multi-passing of the wave due to multiple reflections at interfaces in the case of thin films, as shown Figure 1a. Similarly, the scattered field from depth y must be calculated to take account of any multiple reflections. Anisotropy is accounted for by employing distinct complex optical indices along the zigzag (a) and armchair (c) axes, represented as $N_{1a}$ and $N_{1c}$. From now on, we shall use Jones notations.

The incident field vector is: $\begin{pmatrix} E_{inc.(a)}\cos\theta \\ E_{inc.(c)}\sin\theta \end{pmatrix}$ and the scattered electric field in parallel (in-phase incident and scattered polarizations) and orthogonal (incident and scattered polarizations offset by 90°) configurations are calculated as follows:

$$E_{scat.}^{PARA} = (E_{scat.(a)}\cos\theta \quad E_{scat.(c)}\sin\theta) R \begin{pmatrix} E_{inc.(a)}\cos\theta \\ E_{inc.(c)}\sin\theta \end{pmatrix} \quad (1a)$$

$$E_{scat.}^{ORTHO} = (E_{scat.(a)}\cos\theta \quad E_{scat.(c)}\sin\theta) R \begin{pmatrix} -E_{inc.(a)}\sin\theta \\ E_{inc.(c)}\cos\theta \end{pmatrix} \quad (1b)$$

where θ is the angle between the axis of the incident polarization and the (a) axis of the crystal, and R the Raman tensor. The scattered intensity from the depth y is defined as:

$$I^{PARA}(y) = E_{scat.}^{PARA} \cdot (E_{scat.}^{PARA})^* \text{ in parallel configuration} \quad (2a)$$

$$I^{ORTHO}(y) = E_{scat.}^{ORTHO} \cdot (E_{scat.}^{ORTHO})^* \text{ in orthogonal configuration} \quad (2b)$$

Waves from different depths are treated as incoherent so that the total scattered intensity is the integral over the entire thickness $d_1$ of the BP layer $I = \int_0^{d_1} I(y)dy$.

Detailed expressions and parameters used to calculate the incident and scattered electric fields within the BP/SiO$_2$/Si stack, accounting for anisotropic interference effects, are provided in the Supplementary Information.

### III. Results on bulk BP crystals (thickness > 1 µm)

The BP crystal Raman spectrum is presented Figure 1 and exhibits three characteristic sharp peaks corresponding to vibration modes labelled from their symmetry $A_g^1$ (364 cm$^{-1}$), $B_{2g}$ (441 cm$^{-1}$) and $A_g^2$ (468 cm$^{-1}$)[51]. Figure 2a shows the ARPRS measurements performed for the three modes ($A_g^1$, $B_{2g}$ and $A_g^2$) at different θ angles in parallel (black squares) and orthogonal (white dots) configurations. The polar representations of the modes in the (a,c) plane have characteristic shapes typical of thick BP samples with a bi-lobed $A_g^1$ mode whose maxima are directed along the zigzag axis (a) and a more quadri-lobed $A_g^2$ mode for the parallel configuration[34–36,43]. In this configuration, the $B_{2g}$ maxima point between the crystal axes at 45°, 135°, 225° and 315°.

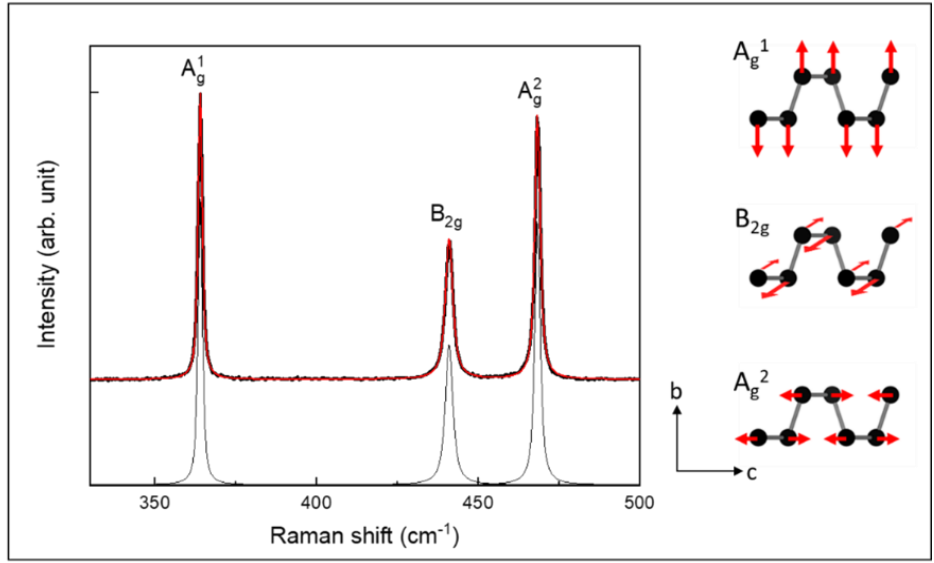

Figure 1. Raman spectrum of the black phosphorus crystal. On the right, the representation in the vibration space of the three observed modes $A_g^1$, $B_{2g}$ and $A_g^2$.

The experimental points are fitted using equations (2a) and (2b), in which the incident and scattered electric fields are calculated using equations detailed in the Supplementary Information and the following Raman tensors:

$$R^{A_g} = \begin{pmatrix} |a|e^{i\varphi_a} & 0 \\ 0 & |c|e^{i\varphi_c} \end{pmatrix} \text{ and } R^{B_{2g}} = \begin{pmatrix} 0 & |f|e^{i\varphi_f} \\ |f|e^{i\varphi_f} & 0 \end{pmatrix} \quad (5)$$

The parameters a, c, f and $\varphi_{ac} = \varphi_a - \varphi_c$ are derived from the Raman tensors of the different phonon modes.

Adjustment results are shown in black line for the parallel configuration and in grey line for the orthogonal configuration in Figure 2a. Only the $A_g$ modes are affected by the birefringence corrections (consideration of different indices depending on light polarization). The $B_{2g}$ mode is not affected because it has only one tensor parameter along both crystallographic directions a and c. We then find the following relationship between the Raman tensor parameters of black phosphorus: $\left|a_{A_g^1}/c_{A_g^1}\right| = 0.53$, $\varphi_{ac_{A_g^1}} = 10°$, $\left|a_{A_g^2}/c_{A_g^2}\right| = 0.32$, $\varphi_{ac_{A_g^2}} = 90°$. They are in good agreement with the values reported in the literature[32,36].

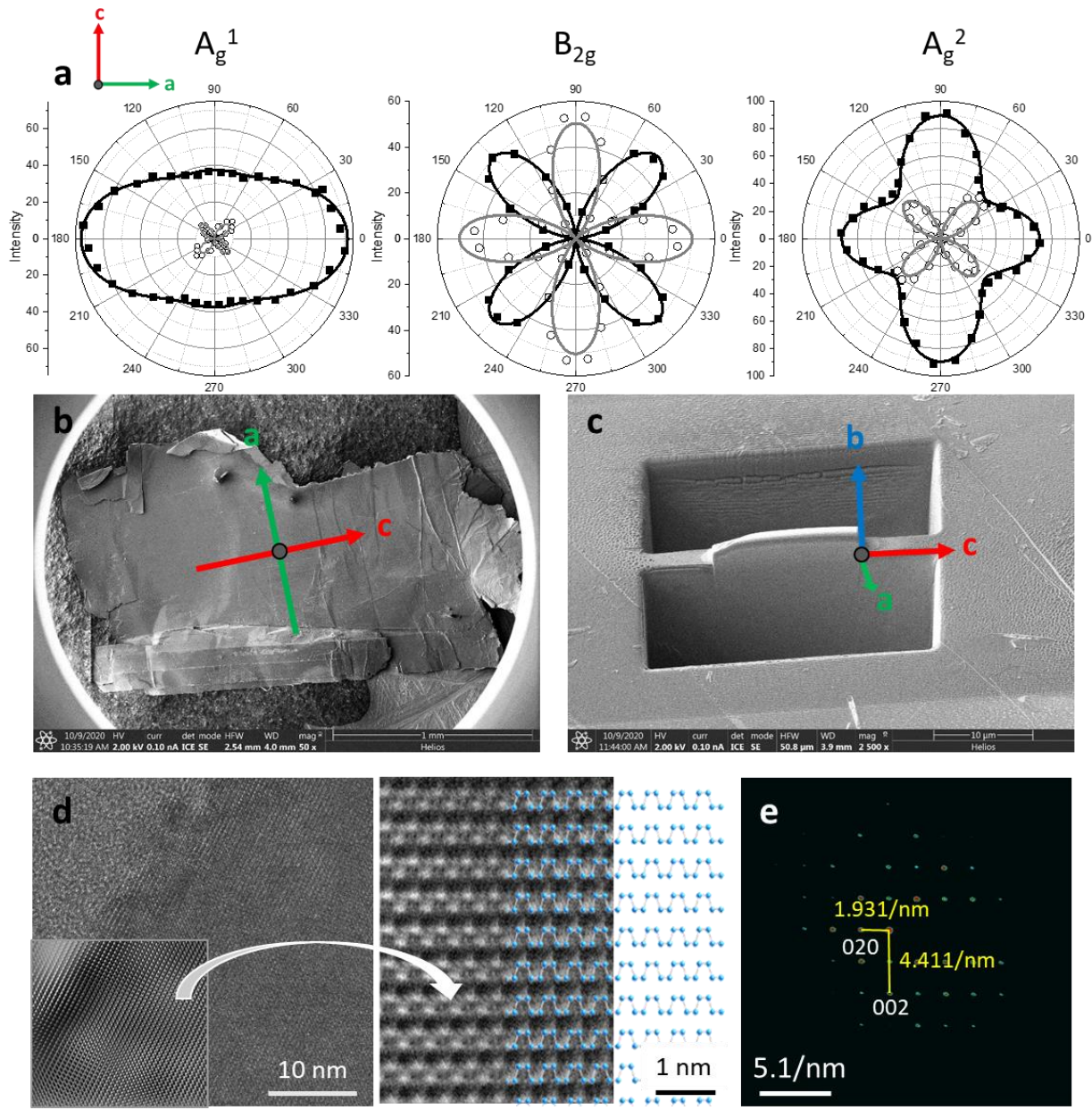

Figure 2. (a) Polar representation of the intensity of Raman modes of a BP bulk crystal (shown in (b)) in parallel (black squares) and orthogonal (white dots) configurations. Fits of experimental data are shown in black (resp. grey) line for parallel (resp. orthogonal) configuration. (b) The crystalline orientation deduced from the Raman measurements is marked on the SEM image of the BP crystal studied by red and green arrows for (c) and (a) axes respectively. (c) SEM image showing the cut of a FIB slide perpendicularly to the (a) axis. (d) HRTEM image of the FIB slide. A Fourier filtering is presented in the inset and a magnified detail is presented on the right with a superimposition of the atomic positions (blue dots) of crystal structure projected along the (a) axis. (e) Diffraction pattern of the FIB slide and its indexation corresponding to the (100) zone axis.

The intensity extinctions of the $B_{2g}$ mode in parallel configuration are reached for the two crystallographic axes. Usually, the extinctions are preferred because they are easily and more precisely identified than intensity maxima. The $B_{2g}$ mode is therefore chosen to define the orientation of the two axes. However, it does not allow them to be discriminated. The full crystallographic orientation is only achieved by examining the orientation of the intensity maxima of the $A_g^1$ mode in parallel configuration which are along the zigzag (a) axis. It is worth mentioning that the analysis in parallel configuration is sufficient for identifying crystal orientation and hereafter only the parallel configuration will be carried out for thin layers.

The present orientation method is only valid for crystal thicknesses greater than the laser's penetration depth. For an excitation at λ=514 nm, we estimate this penetration depth as δ=λ/4πκ, where κ is the extinction coefficient[52]. For the zigzag (a) axis, we find $δ_a$=757 nm and for the armchair (c) axis, $δ_c$=110 nm. Thus, it is expected that the bulk orientation model is valid provided the thicknesses is greater than max($δ_a,δ_c$)=757 nm.

To validate the method, we choose to compare the orientation results given by ARPRS and by a direct TEM measurement. A BP crystal, with a thickness greater than 6 μm (AFM saturation), is oriented using the ARPRS method presented above. Figure 2b shows the SEM image of the crystal as well as the orientation of its crystal axes a and c, extracted from ARPRS measurements, in the SEM stage reference frame. As shown in Figure 2c, a slide is cut by FIB perpendicularly to the zigzag (a) axis so that it is expected to be lying in the (b,c) plane and then examined in TEM. A high-resolution TEM image of the FIB slide is shown in Figure 2d and a Fourier filtering designed to remove noise is presented in the inset. A magnification of the filtered image reveals a dot pattern which is well the one expected for an observation of the crystal structure projected along the a-axis, i.e. in the (100) zone axis. The superposition on the image of the atomic positions in this projection makes it possible to recognize the characteristic "waffle-structure" of BP. Figure 2e shows an electron diffraction pattern of the FIB blade attesting to its orientation along the zone axis (100). The diffraction pattern is very slightly offset from the impact of the incident beam, indicating a very small disorientation from the zone axis of at most 1°. Values of the lattice parameters deduced from this diffraction pattern are $c = 4.5$ Å and $b = 10.4$ Å in good agreement with those established in the literature in X-ray diffraction by Hultgren[53] ($c = 4.38$ Å and $b = 10.50$ Å) and Brown[37] ($c = 4.37$ Å and $b = 10.48$ Å).

In conclusion, the method of orientation of BP massive crystals with ARPRS, based on the examination of extinction of $B_{2g}$ mode and maxima orientation of $A_g^1$ mode in parallel configuration, allowed us to define with precision the crystalline axes and to cut a FIB blade oriented along the a-zone axis. High-resolution TEM and electron diffraction observations of this slide validated the orientation of the cut to better than one degree in angular accuracy and thus validated the reliability of the Raman spectroscopy procedure.

### IV. Investigation of thin BP layers

To complement our study on bulk BP crystals, we provide in the Supporting Information a detailed description of the multiple reflection model used for thin BP layers (Figures SI1 & 2), along with the associated theoretical predictions (Figures SI3). These simulations consider anisotropic interference effects in the BP/SiO₂/Si stack and anisotropic absorption of the BP[54]. As shown by Zou et al.[42], this approach enables the use of intrinsic bulk Raman tensor elements to describe the behavior of thin BP layers down to thicknesses as low as 8 nm. Recent studies[42,55] have indeed demonstrated that the Raman tensor elements remain essentially constant with thickness, owing to the strong similarity in band structure between bulk and few-layer BP, up to the limit where interlayer coupling and quantum confinement effects become significant.

The simulated intensity ratio along the two axes of the plane, shown Figure 3a for $A_g^1$ ($I_{A_g^1}^a/I_{A_g^1}^c$) and Figure 3b for $A_g^2$ ($I_{A_g^2}^a/I_{A_g^2}^c$), reveal three distinct thickness dependent regimes. For thicknesses above ~150 nm, the ratios converge towards the bulk values: the $A_g^1$ mode ratio is consistently greater than 1, indicating a dominant Raman response along the zigzag (a) axis, while the $A_g^2$ mode ratio remains close to 1, reflecting the 4-lobed symmetry observed in bulk samples Figure 2a. Below ~40 nm, both

$A_g$ modes show ratios significantly lower than 1, which indicates that the maximum Raman intensities are aligned along the armchair (c) axis, and confirms the unambiguous orientation of the flakes in the thin regime. Between 40 and 150 nm, both ratios oscillate around unity, resulting in near-isotropic patterns highly sensitive to slight thickness variations, which complicates crystal axis identification. This transitional behavior likely accounts for previously reported inconsistencies in the literature regarding axis assignment in intermediate-thickness BP flakes.

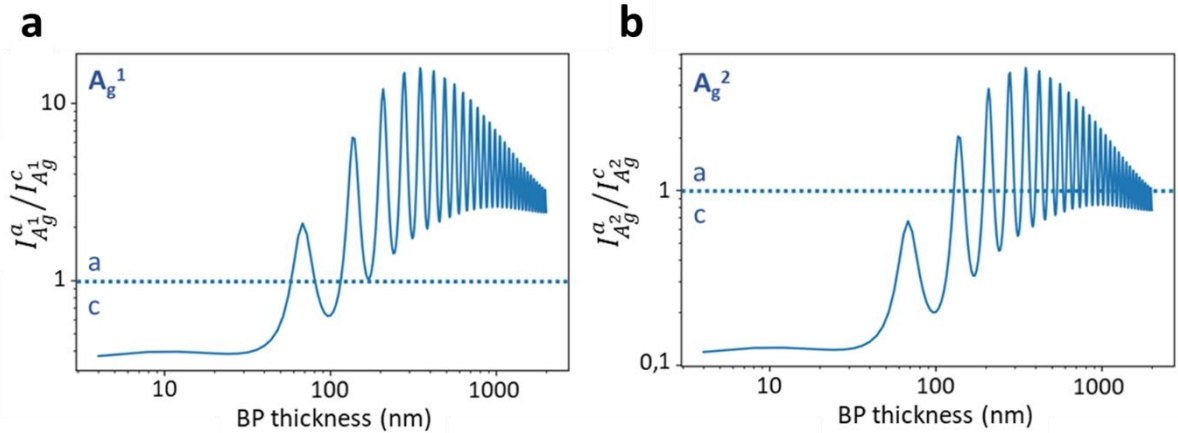

Figure 3. (a) Raman intensity ratios along zigzag and armchair $I^a_{A^1_g}/I^c_{A^1_g}$ for the $A_g^1$ mode and (b) ratio $I^a_{A^2_g}/I^c_{A^2_g}$ for the $A_g^2$ mode. Above (resp. below) the horizontal dotted line $I^a_{A_g}/I^c_{A_g}=1$, the maxima of the $A_g$ modes are directed along the zigzag (resp. armchair) axis.

These theoretical predictions are further confronted to experimental results with EBSD measurements on 2D BP films. Three BP flakes were exfoliated on $SiO_2$(300 nm)/Si substrate and passivated with 1 nm of alumina. SE, EBSD and AFM data of these samples are presented in Figure 4. The thicknesses of the flakes were measured using AFM and found to be equal to 180 nm, 60 nm, and 12 nm for samples 1, 2, and 3 respectively, corresponding to the three thickness regimes identified in Figure 3. The EBSD maps show that the flakes are single crystalline and that their stacking axis (b or 010) is orthogonal to the substrate. The EBSD measurements allow us to find the orientation of the two axes a and c in the xy plane of the plate, as indicated by green and red arrows in the SEM image of each sample. The edges of the flakes do not follow crystallographic directions, unlike more conventional 2D materials with the hexagonal structure, which present easy cleavage directions in the atomic layer plane[56,57]. The "soft" mechanical behavior of BP can be attributed to its heightened interlayer force constant, surpassing other 2D materials such as $MoS_2$ by a factor of five and graphene by ten[58]. Moreover, within its atomic plane, BP exhibits a lower Young's modulus, with values along the armchair axis being eight times weaker than $MoS_2$ and twenty-three times weaker than graphene[28,59].

The ARPRS measurements of the three modes $A_g^1$, $B_{2g}$ and $A_g^2$ for the three samples are shown in Figure 5 where EBSD orientations are reported with colored arrows on their optical images. As for the bulk, the $B_{2g}$ mode allows a partial orientation, by identifying the directions of the perpendicular axes (a) and (c) without discrimination between them. These directions are represented by black arrows in the polar figure of the $B_{2g}$ mode and in the optical images in Figure 5. At this stage, differences in orientation between the ARPRS and EBSD measurements are very small, being 1°, 0° and 4° respectively for samples 1, 2 and 3. The 4° difference is attributed to EBSD orientation uncertainty due to the weak signal of the 12 nm BP layer (sample 3).

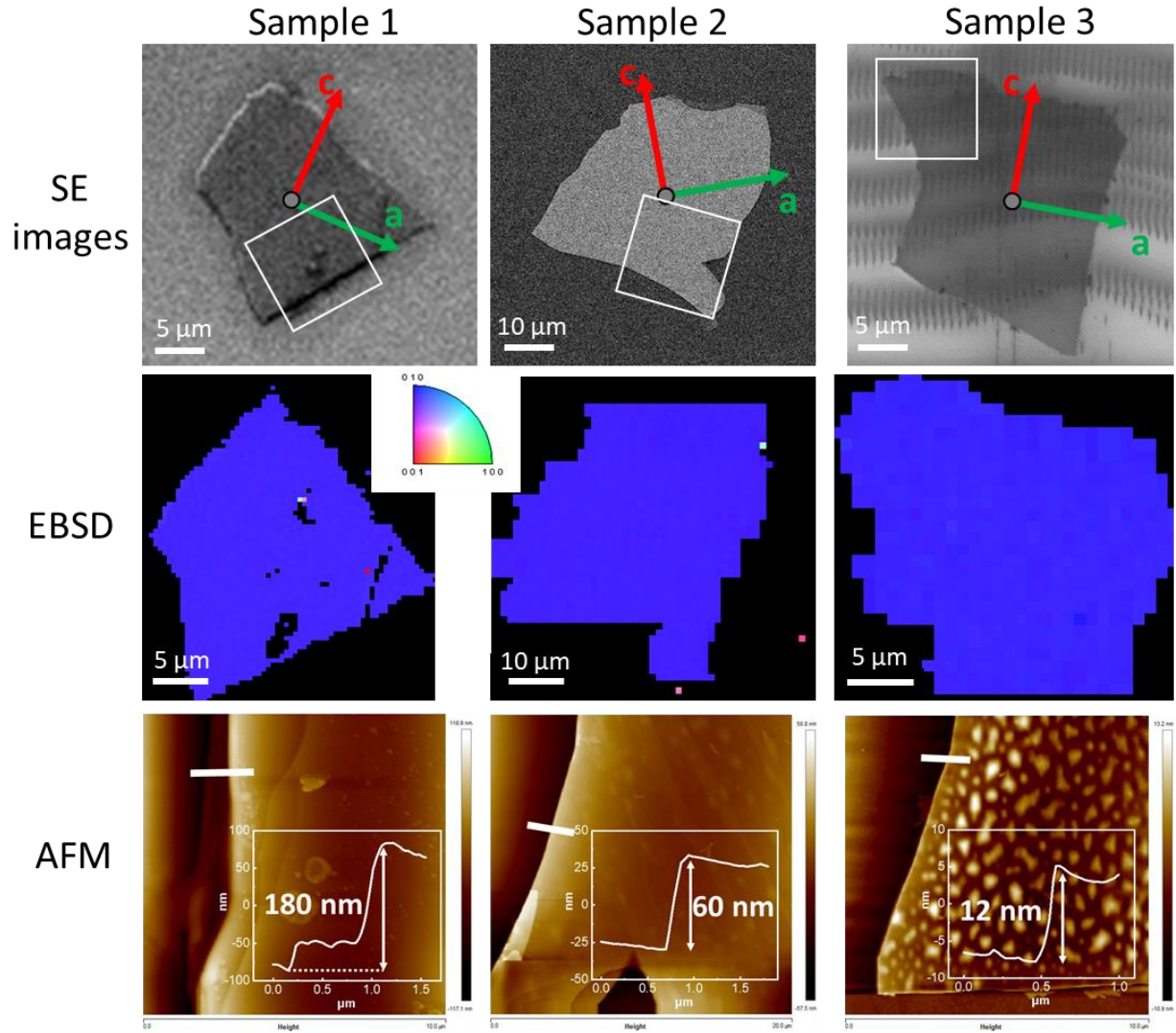

Figure 4. SE images, EBSD mappings and AFM images of the three exfoliated black phosphorus crystals: sample 1 (150 nm), sample 2 (60 nm) and sample 3 (12 nm). The orientation of the crystal lattice deduced from the EBSD measurements is indicated with green and red arrows in the SE images. The area measured in AFM is reproduced by a white square on the SE images.

Figure 5 compares ARPRS measurements related to the $A_g^1$ and $A_g^2$ modes with simulations performed at the thicknesses of the three flakes. For samples 1 and 3, the theory and experimental results show good fair agreement: both $A_g^1$ and $A_g^2$ modes' lobes point to the zigzag crystallographic direction for sample 1 and armchair for sample 3. However, for sample 2, while the $A_g^1$ mode is well described, the theory predicts for $A_g^2$ mode maxima along the armchair axis in contrast to the measurements that show lobes pointing to zigzag axis. Thickness of sample 2 falls in the thickness range for which $A_g^2$ mode is very sensitive to interference effects and probably affected by the strong uncertainties on refractive indexes[36,46,52]. This could also explain the differences in the angular diagram shape for the $A_g^1$ mode where the experimental results show more oblong profiles for the first two samples and more isotropic for the third one compared to the model. These discrepancies suggest that the simulations tend to underestimate the $I_{A_g}^a / I_{A_g}^c$ ratio for both modes. This effect is exacerbated for the thin sample 3 and could be explained by the fact that the refractive indexes[60] and the Raman tensor values[32,43] could evolve at atomic layer thicknesses. As noted by Zou et al.[42], using bulk Raman tensor parameters for BP layers thinner than approximately 10 nm may become less accurate, as the band structure gradually deviates from that of the bulk. Moreover, Luo et al.[41] have identified additional factors that could influence the evolution of these parameters at low thickness, such as reduced thermal conductivity leading to laser-induced heating, faster degradation of BP, and strain-related effects.

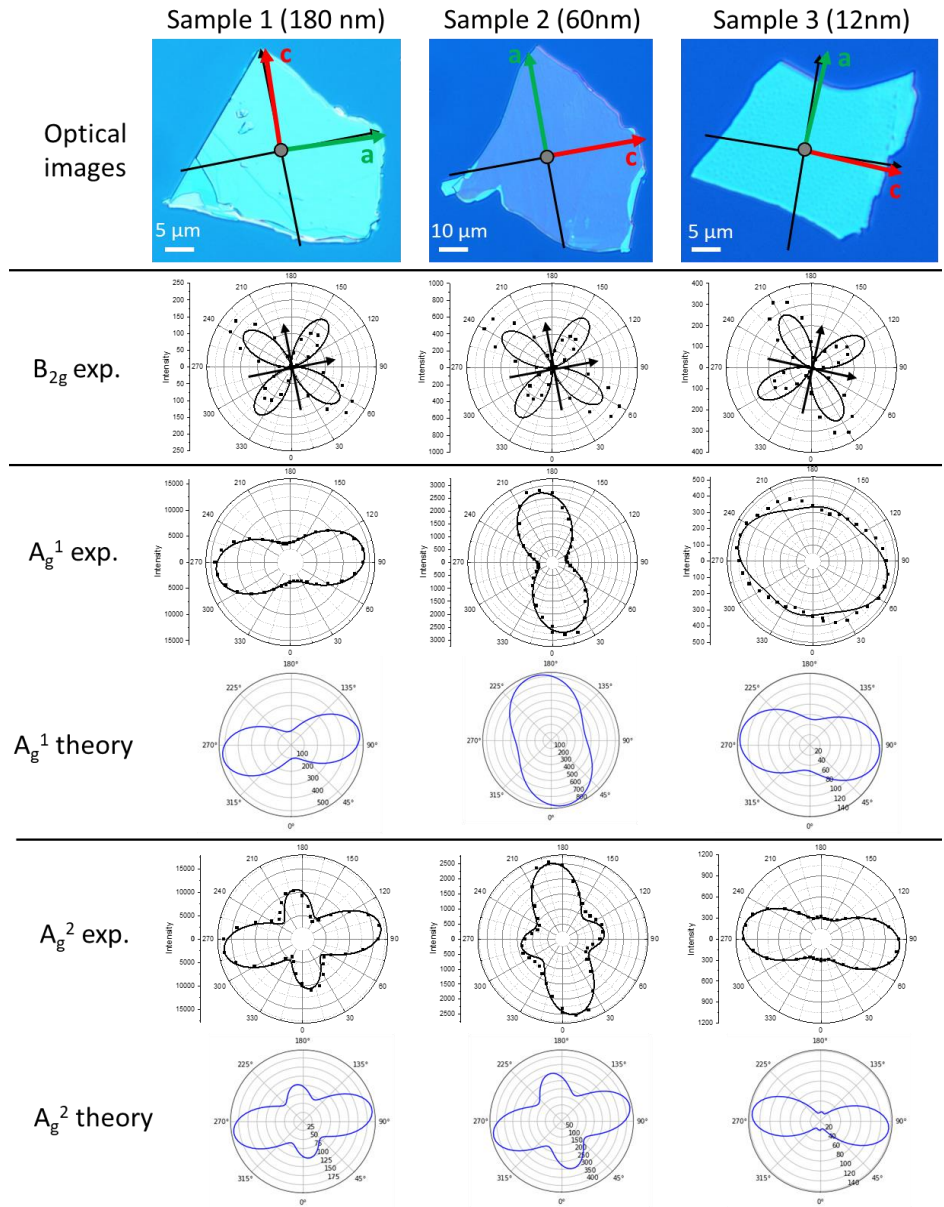

Figure 5. Optical images and polar graphs of the intensities of the $A_g^1$, $B_{2g}$ and $A_g^2$ Raman modes in parallel configuration: experimental (exp. black dots) and simulated from the theory (blue lines) results for samples 1, 2 and 3. The partial crystallographic orientation extracted from measurements of the $B_{2g}$ Raman mode is indicated with black arrows on the optical images. The orientation results by EBSD in Figure 4 are plotted with colored arrows (zigzag-a in green, armchair-c in red) on the optical images. Fits of experimental measurements (grey lines) are performed using eq. 3 to accurately evaluate the angles of maximum and minimum intensities.

As expected from the simulations of Figure 3, we observe an inversion from zigzag to armchair axis for the angular lobe of the $A_g^1$ mode between samples 2 and 3. The $A_g^1$ polar plot points to the zigzag axis in sample 2 and to the armchair axis in sample 3. Given the sample thicknesses, the inversion occurs in the range 12-60 nm. This range is fully consistent with our theoretical prediction that, upon increasing thickness, the first armchair reversal toward zigzag occurs for a thickness of 58 nm. This thickness value is also consistent with experimental data available in the literature. Kim et al.[36]

observed this reversal between 5 and 65 nm (λ = 514 nm), Zou *et al.*[33] between 8 and 57 nm (λ = 532 nm), and Mao *et al.*[46] between 10 and 50 nm (λ = 514 nm). Our experimental and theoretical results confirm that axis identification using $A_g^1$ mode intensities is reliable at both very low thickness (< 60 nm) and high thickness (> 150 nm). In the intermediate range of thickness, a more careful analysis is required.

As recently proposed by Zou *et al.*[33], we confirm that the $I^c_{A_g^2} \cdot I^a_{A_g^1} / I^c_{A_g^1} \cdot I^a_{A_g^2}$ ratio is always strictly larger than 1 for our three samples and according to our theoretical model. This conclusion is directly derived from the ratio of Raman tensor elements in eq. (5). Although these values can change at atomic scale, the ratios appear to be constant from bulk down to relatively low thicknesses (~ 8nm)[33]. Analysis of this ratio could resolve the ambiguity on the axes in the intermediate thickness range where the angular dependence of the $A_g^1$ mode is very sensitive to oscillations of the ratio $I^a_{A_g^1}/I^c_{A_g^1}$ (40 to 150 nm).

## V. Conclusions

In summary, we combined EBSD and TEM techniques to evaluate the effectiveness of an ARPRS protocol for precise BP orientation using a single 514 nm excitation wavelength. First, we validated that measurements of the $A_g^1$ and $B_{2g}$ modes in a parallel configuration are sufficient for a complete orientation of zigzag and armchair axes in bulk crystals. The reliability of this protocol was confirmed by preparing a FIB lamella precisely oriented along the chosen axis for TEM observation.

In analyzing thin films, we addressed substrate interference effects dependent on laser penetration depth, adapting conventional treatments to accommodate BP's anisotropic nature. A comparative ARPRS analysis of samples ranging from 12 to 180 nm, combined with EBSD orientation data and theoretical modeling, allowed us to clarify the constraints of crystal orientation by ARPRS. We observed that, across all thicknesses, the $B_{2g}$ mode consistently provides orientation information for the armchair and zigzag axes, though it does not differentiate between them. For the $A_g^1$ mode, we noted a contrast reversal, where the orientation shifts from zigzag to armchair in layers thinner than 60 nm, compared to layers thicker than 150 nm. This shift is attributed to a marked decrease in the Raman intensity ratio along the zigzag and armchair axes $I^a_{A_g^1}/I^c_{A_g^1}$. In the intermediate thickness range (60–150 nm), contrast shows a strong dependence on thickness due to oscillations in intensity ratios. To accurately distinguish the zigzag and armchair directions within this challenging range, a detailed combined analysis of $A_g^1$ and $A_g^2$ polar plots is required.

In conclusion, this study provides a straightforward method for achieving a full orientation of thin and thick black phosphorus crystals using a simple Raman spectroscopy setup equipped with a single 514 nm laser wavelength. Owing to its non-destructive nature, rapid acquisition, and compatibility with standard equipment, the method holds promise for future adaptation toward high-throughput or in-line characterization workflows, particularly in wafer-scale device fabrication.

## Acknowledgments

The authors acknowledge funding from the French national research agency (ANR) under the grant agreement No. ANR-17-CE24-0023-01 (EPOS-BP). This project has also received funding from the European Union's Horizon 2020 research and innovation program under grand agreement N° 881603 (Graphene Flagship core 3). The authors thank Juba Hamma for his assistance with the python code.